\documentclass[twocolumn,amsmath,amssymb,prb,showpacs]{revtex4}
\usepackage{dcolumn}    
\usepackage{bm}         
\usepackage{graphicx}
\usepackage{amsmath}

\date{\today}
\begin{document}
\title{Hybrid-functional calculations with plane-wave basis sets:
The effect of the singularity correction on total energies,
energy eigenvalues, and defect energy levels}
\author{Peter Broqvist}
\author{Audrius Alkauskas}
\author{Alfredo Pasquarello}
\affiliation{
Institute of Theoretical Physics,
Ecole Polytechnique F\'ed\`erale de Lausanne (EPFL),
CH-1015 Lausanne, Switzerland}
\affiliation{
Institut Romand de Recherche Num\'erique en Physique des Mat\'eriaux (IRRMA),
CH-1015 Lausanne, Switzerland}%

\date{\today}

\begin{abstract}
When described through a plane-wave basis set, the inclusion of
exact nonlocal exchange in hybrid functionals gives
rise to a singularity, which slows down the convergence with
the density of sampled $k$ points in reciprocal space.
In this work, we investigate to what extent the treatment
of the singularity through the use of an auxiliary function
is effective for $k$-point samplings of limited density,
in comparison to analogous calculations performed with
semilocal density functionals. Our analysis applies for instance
to calculations in which the Brillouin zone is sampled at the
sole $\Gamma$-point, as often occurs in the study of surfaces,
interfaces, and defects or in molecular dynamics simulations.
In the adopted formulation, the treatment of the singularity
results in the addition of a correction term to the total energy.
The energy eigenvalue spectrum is affected by a downwards shift of
the energy eigenvalues of the occupied states, while those of the
unoccupied states remain unaffected. Analogous corrections also speed up
the convergence of screened exchange interactions despite the absence
of a proper singularity. Focusing first on neutral systems, both finite
and extended, we show that the account of the singularity corrections
bears convergence properties which are quantitatively similar to
those observed with semilocal density functionals. We emphasize
that this is not the case for uncorrected energies, particularly for
elongated simulation cells for which qualitatively different trends are found.
We then consider differences between total energies of systems differing
by their charge state. For systems involving localized electron states,
such as ionization potentials and electron affinities of molecular systems or
charge transition levels of point defects, the proper account of the
singularity correction yields convergence properties which are similar
to those of neutral systems. In the case of extended systems, such
energy differences provide an alternative way to determine the band edges,
but are found to converge more slowly with simulation cell than in
corresponding semilocal functionals because of the exchange selfinteraction
associated to the extra charge.
\end{abstract}


\maketitle

\section{Introduction}
In the past decades, density functional
theory\cite{Hohenberg_PR_1964,Kohn_PR_1965}
has become the mainstream technique for electronic structure calculations
of large molecules, clusters, liquids, and solids. In condensed matter
applications, the most common functional has initially been based on the
local density approximation (LDA),\cite{Kohn_PR_1965} but generalized gradient
approximations have become increasingly popular in the last two
decades.\cite{Perdew_PRB_1986,Becke_JCP_1986,Becke_PRA_1988,%
Perdew_PRB_1992,Perdew_PRL_1996}
A class of functionals that could potentially lead to higher accuracy
include a fraction of exact nonlocal exchange in the exchange-correlation
potential.\cite{Becke_JCP_1993a,Becke_JCP_1993b}
This class of functionals, referred to as hybrid functionals, has become
the standard electronic-structure approach in quantum-chemistry applications.
For molecular systems, these functionals achieve a more accurate description
not only of atomization energies,\cite{Curtiss_JCP_1997} but also of
ionization potentials and electron affinities.\cite{Curtiss_JCP_1998}
However, several shortcomings still remain. For instance, the desired
chemical accuracy is not always attained and no solution is offered for
the treatment of Van-der-Waals interactions. Nevertheless, it appears quite
clearly that the inclusion of exact exchange constitutes an improvement,
which might be particularly useful in several circumstances.

When applied to semiconductors and insulators, hybrid functionals provide
a superior description to semilocal functionals. For instance,
structural parameters are found to be closer to experimental
values.\cite{Heyd_JCP_2005,Paier_JCP_2006} Furthermore, hybrid
functionals give electronic band gaps which are systematically 
larger than those achieved with semilocal functionals, generally leading to 
a better agreement with experiment.\cite{Muscat_CPL_2001,Heyd_JCP_2005,Paier_JCP_2006}
In particular, the improvement achieved by hybrid 
functionals in which the Coulomb potential is screened is 
remarkable,\cite{Heyd_JCP_2005,Paier_JCP_2006} and the origin of this
successful description can be rationalized.\cite{Janesko_PCCP_2009}   
A more accurate description of the band gap is especially important in
certain classes of problems, such as the study of surface and interface
states\cite{Paier_JCP_2005,DiValentin_PRL_2006} and
the determination of defect energy levels.\cite{VanDeWalle_JAP_2004}
Indeed, a physically meaningful description of electronic states lying in the
band gap can only be achieved when the calculated band gap approaches the
experimental one.\cite{VanDeWalle_JAP_2004,Pacchioni_PRB_2000,Zhang_PRB_2001,Knaup_PRB_2005,%
Deak_JPCM_2005,Xiong_APL_2005,Gavartin_APL_2006,Broqvist_APL_2006,%
DiValentin_PRL_2006,Lany_PRL_2007,Janotti_PRB_2007,Broqvist_APL_2008,Alkauskas_PRL_2008,%
Broqvist_PRB_2008,Paudel_PRB_2008,Oba_PRB_2008,Alkauskas_PRB_2008,Lany_PRB_2008}

For the treatment of condensed systems, hybrid functionals have been available
for some time in codes combining Gaussian basis sets and periodic boundary
conditions.\cite{Pisani_IJQC_1980, Gaussian03}
However, it is expected that the treatment of the electronic structure through
hybrid functionals carries potential to be much more widely used in
an implementation based on plane-wave basis sets and
pseudopotentials.\cite{Payne_RMP_1992}
In this respect, the treatment of exact exchange poses several new problems
with respect to the use of standard semilocal functionals.

First, the calculation of exact exchange entails a significantly higher
computational cost. To address this issue, efficient algorithms have been
developed to optimize the scaling with respect to the number of plane
waves.\cite{Chawla_JCP_1998,Sorouri_JCP_2006,Todorova_JPCB_2006}
A further gain is achieved through optimal adaptation to new massively
parallel computer platforms.\cite{CPMD}
Despite these improvements, plane-wave based hybrid-functional calculations
for a system of 1000 electrons still yield a computational cost
exceeding that of semilocal ones by more than two orders of
magnitude.\cite{Broqvist_APL_2008,Alkauskas_PRL_2008b} Nevertheless,
such calculations are attracting increasing interest as larger
computational resources become available.

Second, the expression for exact exchange includes an integrable
divergence,\cite{Kittel_1963} which hinders its straightforward use within
plane-wave formulations because of its slow convergence with the density
of $k$ points. Gygi and Baldereschi proposed a numerical
treatment of the divergence based on the analytic integration of an
auxiliary function showing the same singularity.\cite{Gygi_PRB_1986}
Such auxiliary functions are nowadays available for arbitrary unit
cells.\cite{Massidda_PRB_1993,Sorouri_JCP_2006,Carrier_PRB_2007}
Alternative treatments consist in truncating the Coulomb
operator,\cite{Spencer_PRB_2008} using a screened Coulomb
potential,\cite{Todorova_JPCB_2006} or transforming the Bloch
functions in order to compute real-space Coulomb integrals.\cite{Wu_PRB_2009}

Third, the nonlocal exchange coupling between valence and core states also
intervenes in the basic pseudopotential approximation. Such interactions
can be accounted for within an all-electron scheme in which the
pseudopotential approximation consists in freezing the wave functions
of the core states.\cite{Paier_JCP_2005} The core-valence
interactions due to exchange then lead to a modification of the nonlocal
pseudopotential term. In consideration of the fact that current hybrid
functionals generally only include a fraction of exact exchange (about 25\%),
pseudopotentials derived within semilocal formulations have often been
transferred to hybrid schemes without any
modification\cite{Todorova_JPCB_2006,Wang_JACS_2007,Broqvist_APL_2008,%
Alkauskas_PRL_2008} or through the only use of nonlinear core
corrections.\cite{Broqvist_PRB_2008} However, when core-valence interactions
are sizable, these practices require particular care, and it has recently
been pointed out that they might lead to
inconsistencies.\cite{Stroppa_NJP_2008}

In this work, we focus specifically on aspects associated with the treatment
of the integrable singularity when calculating the exact exchange expresssion
in electronic-structure schemes based on plane-wave basis sets.
Following Gygi and Baldereschi,\cite{Gygi_PRB_1986} we adopt a formulation
in which the divergence is treated analytically. This scheme can be recast in
such a way that the treatment of the divergence results in a correction term
formally corresponding to the Fourier component of the exchange potential at
vanishing wavevector in the Brillouin zone.\cite{Carrier_PRB_2007}
We first address the degree of convergence that can be achieved with a
$k$-point sampling of finite density for both the total energy and the energy
eigenvalues, in comparison with a similar calculation that does not include
exact exchange. This aspect is particularly important to validate calculations
based on large simulations cells with the Brillouin zone sampled only at the
$\Gamma$ point, a configuration which is often used in surface, interface,
and defect calculations or in {\it ab initio} molecular dynamics simulations.
We also consider exchange
interactions based on screened Coulomb potentials. While this case formally
does not show a singularity, there are specific circumstances in which the
same convergence deficiency occurs as for the bare Coulomb potential.
We then study the effect of the singularity correction on the total energy
of {\it charged} systems. Differences between total energies of systems
involving a different
number of electrons are relevant for determining the electron affinity,
the ionization potential, the band gap, and the charge transition levels of
defects. We discuss the effect of the correction on these quantities for the
cases of both finite and extended systems. In particular, in the case of
extended systems, we clarify the way the singularity correction affects
the band edges obtained from the energy eigenvalues and those obtained
through total-energy differences.

This paper is organized as follows. In Sec.\ \ref{Divergence}, we review
the treatment of the singularity of the exact exchange operator in the
case of plane-wave basis sets following the scheme of Gygi and
Baldereschi.\cite{Gygi_PRB_1986} An extension of this method to the case of
large supercells with sparse $k$-point sampling is described.
A generalized procedure for treating the case of screened exchange is also
given.  In Sec.\ \ref{Methods}, we describe the computational setups used
in this work.  Section \ref{Neutral} presents convergence tests for total
energies, energy eigenvalues, and single-particle energy gaps of
neutral systems including molecules and solids. Results obtained with
$\Gamma$-point sampling are studied for increasing supercell size. In the
case of extended systems, a comparison is carried out with converged
calculations achieved with small unit cells and dense $k$-point samplings.
In Secs.\ \ref{Charge-Loc} and \ref{Charge-Deloc}, we study charged systems
with localized and delocalized charge states, respectively.
In particular, we focus on total energies, single-electron eigenvalues,
electron addition energies, electron removal energies, and charge transition
levels of defects.  We study the convergence of these quantities as a function
of the singularity correction and emphasize the difference between localized
and delocalized states.  The conclusions are drawn in Sec.\ \ref{Conclusions}.

We note that a method related to hybrid functionals consists of including
exact exchange in a local way through the use of an optimized effective
potential.\cite{Kummel_RMP_2008,Stadele_PRB_99} While this scheme is not
explicitly discussed in the following, our considerations regarding the
singularity correction also apply in this case.

\section{Treatment of the singularity} \label{Divergence}
\subsection{Exact exchange} \label{exchange}

The exchange energy of a solid is a finite quantity if expressed per
one unit cell.\cite{Kittel_1963} In a basis of plane waves, the matrix element
of the Fock exchange operator $\hat{V}_{\text{x}}$ is given
by\cite{Gygi_PRB_1986}

\widetext

\begin{eqnarray}
\left\langle \mathbf{k}+\mathbf{G}
\left|\hat{V}_{\text{x}}\right|\mathbf{k}+\mathbf{G}'\right\rangle & = &
-\frac{1}{2\pi^2}\sum_{m,\mathbf{G}''}\int\limits_{\text{BZ}}\! d\mathbf{q}
\frac{c^{*}_{m\mathbf{q}}\left(\mathbf{G}'+\mathbf{G}''\right)c_{m\mathbf{q}}
\left(\mathbf{G}+\mathbf{G}''\right)}{\left|\mathbf{k}-\mathbf{q}-\mathbf{G}''\right|^2}, \label{Fock1}
\end{eqnarray}
where the sum over $m$ runs over the occupied states and the integral is
carried out over the Brillouin zone (BZ).  Since the integrand diverges for
$\mathbf{q}=\mathbf{k}$, special numerical care is required when replacing
the integral with a sum over a finite number of $k$-points.
To treat the singularity, Gygi and Baldereschi used an auxiliary function,
periodic in reciprocal space and showing the same $\sim 1/(\mathbf{k-q})^2$
divergence as the integrand in Eq.~(\ref{Fock1}). The integral of this
function is then subtracted and added to the right hand side of
Eq.~(\ref{Fock1}). The subtracted term eliminates the divergence of the
integrand and turns it into a smooth function of $\mathbf{q}$,
which can accurately be evaluated through a sampling of special $k$-points.
The singularity is effectively transferred to the added term and is taken
care of through analytical integration.\cite{Gygi_PRB_1986,Carrier_PRB_2007,Nguyen_PRB_2009}

The method of Gygi and Baldereschi requires to be adapted in order
to be applied to calculations with large supercells and sparse $k$-point
samplings. First, it is convenient to adopt the notation
$\mathbf{Q}= \mathbf{q}+\mathbf{G}''$ and
$c_{m\mathbf{q}}(\mathbf{G}+\mathbf{G}'')=c_m(\mathbf{G+Q})$:
\begin{eqnarray}
\left\langle \mathbf{k}+\mathbf{G}
\left|\hat{V}_{\text{x}}\right|\mathbf{k}+\mathbf{G}'\right\rangle & = &
-\frac{1}{2\pi^2}\sum_{m}\int\! d\mathbf{Q}
\frac{c^{*}_{m}\left(\mathbf{G}'+\mathbf{Q}\right)c_{m}
\left(\mathbf{G}+\mathbf{Q}\right)}{\left|\mathbf{k}-\mathbf{Q}\right|^2},
\label{FOCKQ}
\end{eqnarray}
where the integral is over the whole of reciprocal space.
For illustration, we focus in the following on a $k$-point sampling
based on the sole $\Gamma$ point, i.e.\ $\mathbf{k}=\Gamma$.
Application of the procedure proposed by Gygi and Baldereschi transforms the right
hand side of Eq.\ (\ref{FOCKQ}) to
\begin{eqnarray}
-\frac{1}{2\pi^2}  \sum_{m} \int d{\mathbf{Q}} \left[
\frac{c^{*}_{m}\left(\mathbf{G}'+\mathbf{Q}\right)c_{m}\left(\mathbf{G}+\mathbf{Q}\right)}
{\mathbf{Q}^2}-c^{*}_{m}\left(\mathbf{G}'\right)c_{m}\left(\mathbf{G}\right) f\left(\mathbf{Q}\right)\right]
-\frac{1}{2\pi^2}\sum_{m}c^{*}_{m}\left(\mathbf{G}'\right)c_{m}\left(\mathbf{G}\right)
\int f\left(\mathbf{Q}\right) d\mathbf{Q},
\label{Fock5}
\end{eqnarray}
where the auxiliary function needs to chosen in such a way that
$f(\mathbf{Q})\rightarrow 1/Q^2$ when $\mathbf{Q} \rightarrow 0$.
In addition, the function $f$ must be integrable in the whole of
reciprocal space.
The term in parentheses is now a smooth function of $\mathbf{Q}$ in which the
divergence is cancelled.
Therefore, it is justified to approximate the integral over
$\mathbf{Q}$ in the first term in Eq.\ (\ref{Fock5}) with a discrete sum via
\begin{equation}
\frac{\Omega}{(2\pi)^2}\int d\mathbf{Q}\rightarrow \sum_{\mathbf{Q}_i\neq0},
\end{equation}
where $\Omega$ is the volume of the simulation cell.  We here assume that
the discretization in $\mathbf{Q}$-space corresponds to the $k$ points for
which the wave functions are determined.
The second term in Eq.\ (\ref{Fock5}) is to be evaluated
through an analytical integration.
The final expression reads:
\begin{eqnarray}
\left\langle \mathbf{G}
\left|\hat{V}_{\text{x}}\right|\mathbf{G}'\right\rangle =
-\frac{4\pi}{\Omega}  \sum_{m} \sum_{\mathbf{Q}_i\neq0} \left[
\frac{c^{*}_{m}\left(\mathbf{G}'+\mathbf{Q}_i\right)c_{m}\left(\mathbf{G}+\mathbf{Q}_i\right)}
{\mathbf{Q}_i^2}-c^{*}_{m}\left(\mathbf{G}'\right)c_{m}\left(\mathbf{G}\right) f\left(\mathbf{Q}_i\right)\right]
-\frac{1}{2\pi^2}\sum_{m}c^{*}_{m}\left(\mathbf{G}'\right)c_{m}\left(\mathbf{G}\right)
\int f\left(\mathbf{Q}\right) d\mathbf{Q}
\nonumber
\label{Fock6}
\end{eqnarray}
In the case of $\Gamma$ point sampling, the sum over ${\mathbf{Q}_i}$ simply corresponds
to a sum over the reciprocal lattice vectors $\mathbf{G}''$ from which $\mathbf{G}''=0$
needs to be excluded.
\endwidetext

We emphasize that the closer $f(\mathbf{Q})$ is to $1/\mathbf{Q}^2$, the
smoother the integrand in the parentheses, and the more accurate the
approximation of the integral by a sum of discrete terms. Hence, this
condition should guide the choice of the optimal function $f$. We note
that in the formulation of Gygi and Baldereschi, the auxiliary function
is to a certain extent arbitrary, because ultimate convergence can always
be achieved by increasing the density of the $k$-point sampling. In the
case of a fixed $k$-point sampling, the converged result is achieved by
considering simulation cells of increasing size.  However, in practice, 
this limit is computationally more prohibitive. For instance, as a word 
of caution, we emphasize that for small band-gap materials very large 
supercells might be required in calculations relying only on the 
$\Gamma$-point even at the semilocal level. Generally, a satisfactory
target for calculations including exact exchange corresponds to the 
achievement of the same level of convergence as that attained for a 
semilocal functional under the same $k$-point sampling conditions. 
For this reason, it is important to chose a function $f$ which yields 
the best convergence properties for the chosen $k$-point sampling.

By reorganizing the order of the terms in Eq.\ (\ref{Fock6}), we obtain the
following appealing form:
\begin{eqnarray}
&&\left\langle \mathbf{G}
\left|\hat{V}_{\text{x}}\right|\mathbf{G}'\right\rangle = \nonumber \\
&&-\sum_{m,\mathbf{G}''}
c^{*}_{m}\left(\mathbf{G}'+\mathbf{G}''\right)c_{m}
\left(\mathbf{G}+\mathbf{G}''\right)
\Phi\left(\mathbf{G}''\right),
\label{Fock7}
\end{eqnarray}
where $\Phi\left(\mathbf{G}\right)$ represents a suitable generalization of
the Fourier transform of the exchange potential and is given by:
\begin{eqnarray}
\Phi\left(\mathbf{G}\right) = \left\{ \begin{array}{ll}
\displaystyle \frac{1}{\Omega}\frac{4\pi}{G^2} & \textrm{for } \mathbf{G} \neq 0,\\
 \chi & \textrm{for } \mathbf{G} = 0,
\end{array} \right.
\label{Fourrier}
\end{eqnarray}
with
\begin{equation}
\chi= \frac{1}{2\pi^2}\int\limits_{\text{All}} f\left(\mathbf{Q}\right)d
\mathbf{Q}-\frac{4\pi}{\Omega}\sum_{\mathbf{G}\neq0} f\left(\mathbf{G}\right).
\label{beta1}
\end{equation}
This form is particularly convenient since it implies that only the
$\mathbf{G}=0$ term needs to be modified in practical implementations.
We note that standard numerical implementations do not experience any
difficulty for treating the differences between large numbers that this
reorganization of terms implies.

\begin{figure}
\includegraphics[width=7.5cm]{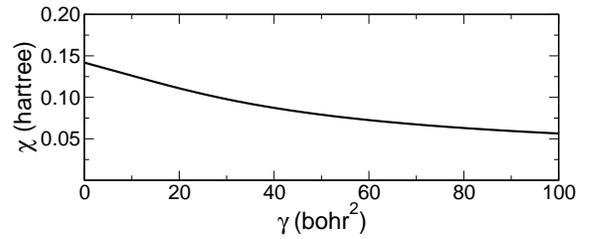}
\caption{The function $\chi$ [Eq.\ (\ref{beta2}] vs the parameter $\gamma$
for a cubic simulation cell with a side of 20 bohr and $\Gamma$-point
sampling. The optimal singularity correction is achieved for $\gamma=0$.}
\label{fig1}
\end{figure}

A suitable form of the auxiliary function $f$
is:\cite{Massidda_PRB_1993,Sorouri_JCP_2006}
\begin{equation}
f\left(\mathbf{Q}\right)=\frac{e^{-\gamma Q^2}}{Q^2} \label{F}
\end{equation}
For this particular choice of auxiliary function, the $\mathbf{G}$=$0$ term
of the exchange potential becomes:
\begin{equation}
\chi\left(\gamma\right)= \frac{1}{\sqrt{\pi \gamma}} -
\frac{4\pi}{\Omega}\sum_{\mathbf{G}}\frac{e^{-\gamma G^2}}{G^2}.
\label{beta2}
\end{equation}
For illustration, we show in Fig.\ \ref{fig1} the dependence of
$\chi$ on $\gamma$ in the case of a cubic cell with a side of 20 bohr.
As mentioned above, the approximation of replacing the integral
in Eq.~(\ref{Fock5}) with a discrete sum is the more accurate, the
closer $f(\mathbf{Q})$ is to $1/\mathbf{Q}^2$.
This leads to the following well-defined expression for $\chi$:
\begin{equation}
\chi = \chi(0) = \lim_{\gamma\rightarrow 0}\left[\frac{1}{\sqrt{\pi \gamma}} -
\frac{4\pi}{\Omega}\sum_{\mathbf{G}}\frac{e^{-\gamma G^2}}{G^2}\right].
\label{beta3}
\end{equation}

We now analyze the effect of the singularity correction on eigenvalues
and total energies.  In our notation, the superscript ``$\text{corr}$''
indicates that the correction $\chi$ has been accounted for, whereas
the superscript ``$\text{uncorr}$'' is used for quantities
obtained with $\Phi(\mathbf{G}$=$0)=0$ in the definition (\ref{Fourrier}).

The exact exchange term contributes to the eigenvalue $\varepsilon_n$ of
the state $n$ through the following expression:
\begin{equation}
\Delta \varepsilon_n = \sum_{\mathbf{G},\mathbf{G'}}c^{*}_{n}\left(\mathbf{G}\right)
c_{n}\left(\mathbf{G'}\right)\left\langle\mathbf{G}\left|\hat{V}_{\text{x}}\right|\mathbf{G'}\right\rangle,
\end{equation}
where we used the matrix element
$\left\langle\mathbf{G}\left|\hat{V}_{\text{x}}\right|\mathbf{G'}\right\rangle $
given in Eq.~(\ref{Fock7}). The singularity correction affects the
eigenvalue as follows:
\begin{eqnarray}
&& \Delta \varepsilon_n^\text{corr} - \Delta \varepsilon_n^\text{uncorr}= \nonumber \\
&& =-\chi\sum_{\mathbf{G},\mathbf{G'}}\sum_{m}c^{*}_{n}\left(\mathbf{G}\right)c_{n}
\left(\mathbf{G'}\right)
c^{*}_{m}\left(\mathbf{G}\right)c_{m}\left(\mathbf{G'}\right) \nonumber\\
&& =-\chi\sum_{m}\delta^2_{nm} = \left\{
\begin{array}{ll}
   -\chi & \text{for {\it n} occupied,}\\
    0     & \text{for {\it n} unoccupied,}
\end{array} \right.
\end{eqnarray}
where $m$ runs over the occupied states and where we used the
orthonormalization condition of the eigenstates.  Thus, all the
eigenvalues of occupied states shift down by $\chi$, whereas
those of unoccupied states remain unaffected:
\begin{equation}
\begin{array}{lcl}
\varepsilon_{n}^{\text{corr}}=\varepsilon_{n}^{\text{uncorr}}-\chi && \text{for {\it n} occupied},\\
\varepsilon_{n}^{\text{corr}}=\varepsilon_{n}^{\text{uncorr}}      && \text{for {\it n} unoccupied}.
\end{array}
\label{corr_eigen}
\end{equation}
In this derivation, we assumed that the ordering of the states is not
affected by the application of the singularity correction.

The exact exchange part of the total energy is given by
\begin{equation}
E_{\text{x}} = \frac{1}{2}\sum_{n}\sum_{\mathbf{G},\mathbf{G'}}c^{*}_{n}\left(\mathbf{G}\right)
c_{n}\left(\mathbf{G'}\right)\left\langle\mathbf{G}\left|\hat{V}_{\text{x}}\right|\mathbf{G'}\right\rangle,
\end{equation}
where the contribution of the singularity correction is
\begin{equation}
-\frac{\chi}{2}\sum_{\mathbf{G},\mathbf{G'}}\sum_{n,m}c^{*}_{n}\left(\mathbf{G}\right)c_{n}
\left(\mathbf{G'}\right)
c^{*}_{m}\left(\mathbf{G}\right)c_{m}\left(\mathbf{G'}\right) = -\frac{\chi N_{\text{el}}}{2},
\end{equation}
with $N_\text{el}$ corresponding to the number of electrons in the supercell.
Thus:
\begin{equation}
E_{\text{tot}}^{\text{corr}}=E_{\text{tot}}^{\text{uncorr}}-\frac{\chi N_{\text{el}}}{2},
\label{corr_en}
\end{equation}
i.e.\ a correction of $-\chi/2$ for each electron.

In the limit $\gamma\rightarrow 0$ [cf.\ Eq.\ (\ref{beta3})], the
correction $-\chi/2$ corresponds to the electrostatic energy of a point
charge interacting with a uniform compensating background charge in the
periodic cell. Indeed, following the method proposed by
Ewald,\cite{Ewald_AP_1921} one replaces the point charge by a Gaussian
charge distribution,
\begin{equation}
\rho(\mathbf{r})=\frac{1}{8\gamma^3\pi^{3/2}}e^{-r^2/(2\gamma)^2}.
\end{equation}
The second term in Eq.~(\ref{beta2}) then corresponds to the electrostatic
energy of the Gaussian charge distribution interacting with the background
and is evaluated in Fourier space. The first term in Eq.~(\ref{beta2})
compensates for the selfinteraction of the Gaussian charge. The complete
Ewald method also gives a third term, which accounts for the difference
between the point charge and the Gaussian charge and which is generally
evaluated in real space.  The absence of this term in Eq.~(\ref{beta2})
accounts for the variation of $\chi$ with $\gamma$ in Fig.\ \ref{fig1}.
However, in the limit $\gamma\rightarrow 0$, the latter term vanishes and
the correction $-\chi/2$ precisely corresponds to the electrostatic energy
of the point charge.  This connection has been pointed out earlier in the
case of isolated molecules in large supercells on the basis of an intuitive
reasoning.\cite{Paier_JCP_2005} The present derivation shows that the same
correction term also applies to the case of extended systems.

\subsection{Screened exchange}

We note that the methodology described above does not only apply in the
presence of a divergence, but might also be useful when the interaction
potential only shows a rapidly varying behavior. Indeed, the direct treatment
of such a potential in Fourier space is difficult when the density of $k$
points cannot easily be increased.  For instance, this applies to screened
exchange interactions.

We here illustrate this point by focusing on the screened exchange interaction
recently proposed in functionals developed by Heyd, Scuseria, and Ernzerhof
(HSE),\cite{HSE03} but the scheme also applies analogously to other forms of
screened exchange.  In the HSE functional, a complementary error function
is used to describe the short-range exchange interaction:
\begin{equation}
V_{\text{sr}}=\frac{\text{erfc}(\omega r)}{r},
\label{sr}
\end{equation}
where $\omega$ is a suitable parameter defining the extent of the potential.
For simplicity, let us again consider the $\Gamma$-point approximation.
The matrix element of the screened exchange operator in a plane-wave basis
set is given by an expression analogous to Eq.~(\ref{Fock7}). The interaction
potential is given by the Fourier transform of the potential defined
in Eq.\ (\ref{sr}), i.e.\
\begin{equation}
\displaystyle{
\Phi_{\text{sr}}\left(\mathbf{G}\right)=\frac{1}{\Omega}\frac{4\pi}{G^2}\displaystyle\left[1-e^{-G^2/(2\omega)^2}\right]}.
\end{equation}
The $\mathbf{G}$=$0$ component of this potential
$\Phi_{\text{sr}}\left(\mathbf{G}=0\right)$ is not divergent, and is equal
to $\pi/(\Omega\omega^2)$.  However, the latter expression cannot be used
for any value of $\omega$. Indeed, when the screened exchange interaction
approaches the exact one (viz.\ in the limit $\omega\rightarrow 0$),
$\Phi_{\text{sr}}\left(\mathbf{G}=0\right)$ diverges rather than converging
to the correct value given in Eq.~(\ref{beta3}).

We derive a suitable expression of $\Phi_{\text{sr}}\left(\mathbf{G}=0\right)$
by proceeding in the same way as in the previous section. The correction
is again given by Eq.~(\ref{beta1}), where we now choose as auxiliary
function:
\begin{eqnarray}
f\left(\mathbf{Q}\right)&=&
\frac{e^{-\gamma Q^2}}{Q^2}\left[1-e^{-{Q^2}/(2\omega)^2}\right] \nonumber \\ \label{F2}
&=&\frac{e^{-\gamma Q^2}}{Q^2} - \frac{e^{-\left(\gamma+\frac{1}{4\omega^2}\right) Q^2}}{Q^2}.
\end{eqnarray}
Using the definition given in Eq.~(\ref{beta2}) and taking the limit as in
Eq.~(\ref{beta3}), we find the correct expression for the $\mathbf{G}$=$0$
component of the screened exchange potential:
\begin{equation}
\Phi_{\text{sr}}\left(\mathbf{G}=0\right)
= \chi(0)-\chi\left(\frac{1}{4\omega^2}\right)= \tilde{\chi}(\omega),
\label{beta_sr}
\end{equation}
which defines the function $\tilde{\chi}(\omega)$.
This leads us to the following form for the interaction potential:
\begin{eqnarray}
\Phi_{\text{sr}}\left(\mathbf{G}\right) = \left\{ \begin{array}{ll}
\displaystyle
\frac{1}{\Omega}\frac{4\pi}{G^2}\displaystyle\left[1-e^{-{G^2}/(2\omega)^2}\right] & \textrm{for } \mathbf{G} \neq 0,\\ \\
 \tilde{\chi}(\omega)                                                                   & \textrm{for } \mathbf{G} = 0.
\end{array} \right.
\label{Fourrier2}
\end{eqnarray}

\begin{figure}
\includegraphics[width=7.5cm]{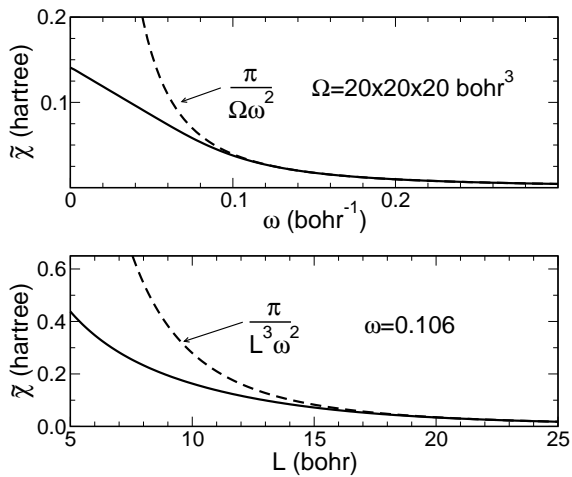}
\caption{(a) The singularity correction for screened exchange
$\tilde{\chi}$ [Eq.\ (\ref{beta_sr})] as a function of the screening parameter
$\omega$ for a cubic simulation cell of 20 bohr (solid line). At large $\omega$ the
singularity correction coincides with the analytical expression
$\tilde{\chi}(\omega)\rightarrow\pi/(\Omega\omega^2)$ (dashed line).
(b) $\tilde{\chi}$ vs the side $L$ of the cubic simulation cell at fixed
$\omega=0.106$ bohr$^{-1}$ (solid line), corresponding to the value set in the
Heyd-Scuseria-Ernzerhof functional (Ref.\ \onlinecite{HSE06}).
At large $L$, $\tilde{\chi}$ approaches $\pi/(L^3\omega^2)$
(dashed line).  The Brillouin zone is sampled at the $\Gamma$ point.}
\label{fig2}
\end{figure}

To illustrate the behavior of this correction as a function of $\omega$,
we considered a cubic simulation cell with a side of 20 bohr.
Figure \ref{fig2}(a) shows that the function $\tilde{\chi}(\omega)$
is nearly indistinguishable from the analytical expression
$\pi/(\Omega\omega^2)$ when $\omega>0.1$ bohr$^{-1}$. However, for lower
values of $\omega$, the screened potential approaches the bare Coulomb
interaction and the analytical expression gives erroneous results.
Instead, the function $\tilde{\chi}(\omega)$ correctly reproduces the
Coulomb limit for $\omega=0$.

For comparison, the parameter $\omega$ assumes the value of 0.106 bohr$^{-1}$
in the HSE functional.\cite{HSE06} Hence, for the case illustrated
in Fig.\ \ref{fig2}(a),
the proposed treatment would not produce a sizable correction.
However, the situation changes when smaller simulation cells are used.
We show in Fig.\ \ref{fig2} the dependence of the singularity correction on
the size of the cubic cell for $\omega=0.106$ bohr$^{-1}$.
It is seen that the analytical expression deviates from the proper
correction $\tilde{\chi}$ for cell sizes smaller than 18 bohr.
This behavior reflects the fact that at these cell sizes the $\Gamma$-point
sampling in reciprocal space is too sparse for properly treating the spatial
varations of the screened exchange potential defined by
$\omega=0.106$ bohr$^{-1}$.

\section{Methods \label{Methods}}

The semilocal density-functional calculations in this work were performed
within the generalized gradient approximation proposed by Perdew, Burke,
and Ernzerhof (PBE).\cite{Perdew_PRL_1996} We considered the class of hybrid
functionals which are obtained by replacing a fraction $\alpha$ of the
PBE exchange with exact exchange:\cite{Perdew_JCP_1996}
\begin{equation}
E_{\text{x}}^{\text{hybrid}}=\alpha E_{\text{x}}^{\text{exact}}+\left(1-\alpha\right)
E_{\text{x}}^{\text{PBE}}.
\end{equation}
In this work, we used the functional defined by $\alpha=0.25$, which
is referred to as PBE0.\cite{Perdew_JCP_1996}
The singularity correction then corresponds to a fraction $\alpha$ of the
correction pertaining to full exact exchange:
\begin{equation}
\beta=\alpha \chi=0.25\chi.
\end{equation}

In our electronic structure scheme based on plane-waves, only valence electrons
are treated explicitly and core-valence interactions are described by
normconserving pseudopotentials.\cite{Troullier_PRB_1991,Goedecker_PRB_1996}
Pseudopotentials were generated at the PBE level and were also used
in the calculations based on hybrid functionals. This is expected to
be a valid approximation for the atoms considered in this work,
i.e.\ C, N, O, and Si, in which core-valence exchange interactions are weak.
We used a kinetic energy cutoff of 20 Ry for systems involving
only Si atoms, but increased the cutoff to 70 Ry when any of the other
atoms occurred.  These cutoffs are sufficiently high to ensure converged
total energies and energy eigenvalues.
For all calculations involving small supercells and dense $k$-point meshes, we
used the code \textsf{PWSCF} of the \textsf{Quantum-ESPRESSO}
package,\cite{quantum-espresso} in which exact exchange and the singularity
correction are implemented.\cite{Nguyen_PRB_2009} All calculations involving 
large supercells and $\Gamma$-point samplings were carried out with the \textsf{CPMD} code.
In this code, we extended the available implementation of exact exchange to
account for the singularity following the scheme outlined in
Sec.\ \ref{exchange}.

We also performed electronic structure calculations for molecules
with the \textsf{Gaussian03} suite of programs.\cite{Gaussian03} We used
the large cc-pVTZ basis set. This method does not give rise to any
divergence of the interaction potential and allowed us to
obtain reliable benchmarks.

\section{Neutral systems} \label{Neutral}

\subsection{Finite systems} \label{Molecules}

\begin{figure}
\includegraphics[width=8.5cm]{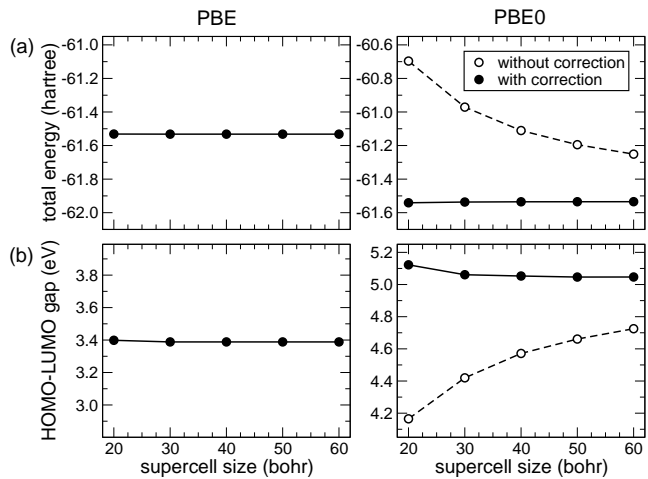}
\caption{(a) Total energies and (b) energy eigenvalue gaps of naphthalene
calculated using the PBE (left panels) and PBE0 (right
panels) functionals vs supercell size. To allow a fair
comparison, the same energy scales are used in the left
and right panels. For PBE0, closed and open symbols indicate values
obtained with the singularity correction turned on and
off, respectively.
}
\label{fig3}
\end{figure}

In this section, we present calculations for total energies and energy
eigenvalues of small organic molecules. For these calculations, we used the
electronic structure scheme based on plane waves and pseudopotentials with
large supercells of varying size and $\Gamma$-point sampling.
We considered the molecules of naphthalene (C$_{10}$H$_8$) and pyridine
(C$_5$H$_5$N), which are sufficiently small to allow us to investigate
the asymptotic behavior for large simulation cells. Furthermore,
the eigenvalue of the lowest unoccupied molecular orbital (LUMO)
of these molecules is found below the vacuum level,
both in PBE and PBE0. In Fig.~\ref{fig3}, the total energies and the
energy eigenvalue gaps of naphthalene calculated in PBE and PBE0
are plotted as a function of the size of the cubic simulation cell.
In hybrid functional calculations, we explicitly compare results
obtained with and without the singularity correction $\beta$, i.e.\
obtained by turning on and off the $\mathbf{G}=0$ component of the exact
exchange potential. One notices that both the total energies and
the energy eigenvalue gaps calculated in PBE0 show a convergence
behavior similar to that achieved in PBE provided the singularity is
accounted for. Note that for the largest considered supercell
(side of 60 bohr) the singularity correction $\beta$ is still
quite sizable and amounts to about 0.2 eV.
A similar behavior is observed for pyridine (not shown).
These results therefore indicate that in the case of isolated molecular
systems, the singularity correction is crucial to achieve well-converged
values of total energies and energy eigenvalues in hybrid functional
calculations based on plane-wave basis sets.

\begin{table}
\caption{Energy eigenvalues of the highest occupied molecular orbitals
($\varepsilon_{\text{HOMO}}$) and of the lowest unoccupied
molecular orbitals ($\varepsilon_{\text{LUMO}}$), energy eigenvalue band gaps
($E_\text{g}$), and ionization potentials (IP)
of naphthalene and pyridine calculated with the PBE0 functional through
a plane-wave ({\sf CPMD}, Ref.\ \onlinecite{CPMD}) and an all-electron
scheme ({\sf Gaussian03}, Ref.\ \onlinecite{Gaussian03}).
The eigenvalues are referred to the vacuum level.}
\begin{ruledtabular}
\begin{tabular}{llrr}
    &                        &    \textsf{CPMD}    &   \textsf{Gaussian03}  \\
\hline
naphthalene  &  $\varepsilon_{\text{HOMO}}$   &   $-$6.34    &   $-$6.28   \\
             &  $\varepsilon_{\text{LUMO}}$   &   $-$1.29    &   $-$1.20   \\
             &  $E_\text{g}$                  &    5.05      &    5.08     \\
             &  IP                            &    7.96      &    7.94     \\
\hline
pyridine     &  $\varepsilon_{\text{HOMO}}$   &   $-$7.51    &   $-$7.43    \\
             &  $\varepsilon_{\text{LUMO}}$   &   $-$0.98    &   $-$0.83    \\
             &  $E_{\text{g}}$                &    6.53      &    6.60      \\
             &  IP                            &    9.51      &    9.46      \\
\end{tabular}
\end{ruledtabular}
\label{tab1}
\end{table}

To benchmark our hybrid-functional results, we also used an all-electron
electronic-structure scheme based on localized atomic orbitals and open
boundary conditions.\cite{Gaussian03} Total energies cannot be compared
because of the different number of electrons that are treated explicitly.
The energy eigenvalues and energy gaps are compared with those obtained
with the plane-wave scheme in Table \ref{tab1}. The eigenvalues
are referred to the vacuum level corresponding to local potential
far from the molecule. The agreement between the two sets of calculations
is very good. This agreement further supports the validity of the
singularity correction derived in Sec.\ \ref{Divergence}. To the extent
that basis set errors for these molecules are small, the differences also
provide an estimate of the way the different treatments of core electrons
affect the energy eigenvalues obtained with hybrid functionals. Table
\ref{tab1} also contains calculated ionization potentials, of
which the discussion is deferred to Sec.\ \ref{finite}.

\subsection{Extended systems} \label{Bulk}

In this section, we study the effect of the singularity correction in
hybrid functional calculations on the total energies and energy
eigenvalues of bulk systems.  In particular, our purpose is
{\it (i)} to illustrate the convergence of small-cell calculations
with $k$-point sampling (cf.\ Refs.\
\onlinecite{Carrier_PRB_2007,Spencer_PRB_2008}) and
{\it (ii)} to study the convergence of $\Gamma$-point calculations
with supercell size. In the latter case, one issue of interest is
whether the achieved level of convergence is similar to that
of corresponding calculations with semilocal functionals.

We chose silicon and $\alpha$-quartz SiO$_2$ to examine systems with
different band gaps. To allow a comparison between PBE and PBE0 calculations
of the electronic structure, we used the lattice parameters optimized
in the PBE also in the hybrid functional calculations.

\begin{figure}
\includegraphics[width=8.5cm]{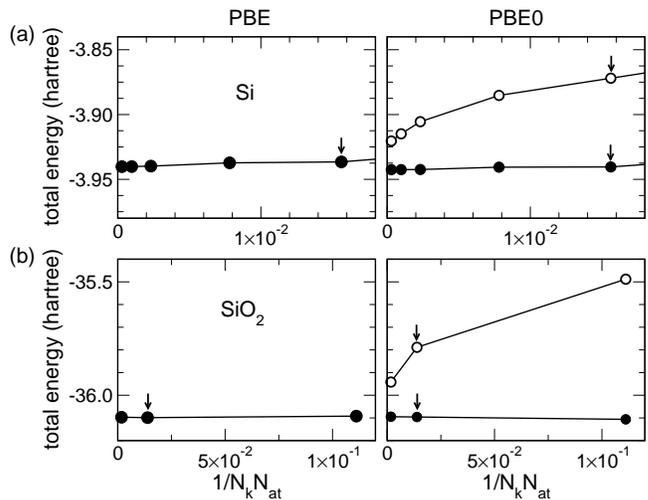}
\caption{\label{fig4} Total energies of (a) Si and (b) $\alpha$-quartz SiO$_2$
per formula unit versus 1/$N_{k}N_\text{at}$, where $N_{k}$
is the total number of $k$-points and $N_{\text{at}}$ the total number
of atoms in the supercell. Results obtained in the PBE and the PBE0
are reported in left and right panels, respectively.
For PBE0, closed and open symbols indicate values obtained with the
singularity correction turned on and off, respectively.
Arrows show data points which were also obtained
with $\Gamma$-point sampling.}
\end{figure}

We first considered the convergence of total energies. PBE and PBE0 results
for silicon and $\alpha$-quartz are displayed in Fig.~\ref{fig4}.
The data points illustrate how convergence is achieved with increasing
density of $k$-point sampling. For both systems, the inclusion of the
singularity correction in the hybrid functional calculations leads to a
faster convergence, very similar to that achieved by the semilocal functional.
It is apparent that the singularity correction is sizable and that its
inclusion significantly speeds up the convergence.
For comparison, we also calculated total energies employing large
supercells and $\Gamma$-point sampling. In the respective limits
of dense $k$-point samplings and large supercell size, the two kinds of
calculations give the same converged energies. When the energies calculated
in the latter scheme are reported in Fig.\ \ref{fig4} (see arrows), they
are found to correspond to energies obtained with
primitive cells and dense $k$-point samplings. In particular, for finite
supercell size, the degree of convergence achieved with hybrid functionals
when the singularity is treated is comparable to that obtained with semilocal
functionals. These results highlight the importance of including the
singularity correction when using hybrid functionals with $\Gamma$-point
sampling.  This is particularly important when total energies obtained with
supercells of different size are comparatively evaluated, such as in the
optimization of lattice parameters or in constant-pressure
molecular dynamics simulations.

\begin{figure}
\includegraphics[width=8.5cm]{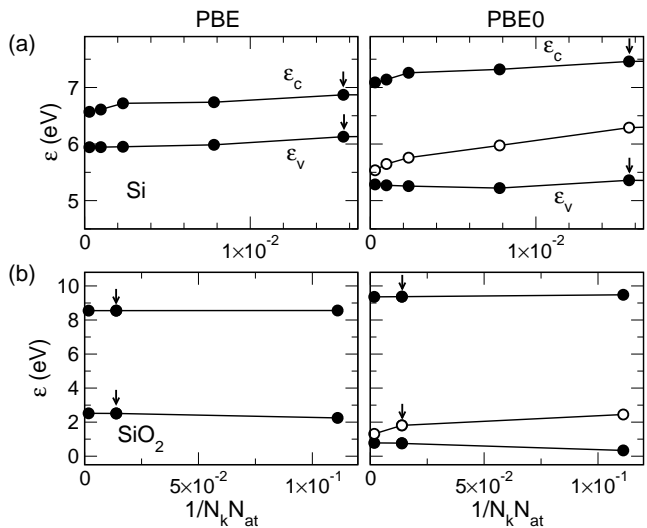}
\caption{\label{fig5}
Eigenvalues corresponding to the valence ($\varepsilon_{\rm v}$) and
conduction ($\varepsilon_{\rm c}$) band edges
of (a) Si and (b) $\alpha$-quartz SiO$_2$ vs
1/($N_{k}$$N_{\text{at}}$), where $N_{k}$ is the total number of $k$-points
and $N_{\text{at}}$ the total number of atoms in the supercell.
Results obtained in the PBE and the PBE0
are reported in left and right panels, respectively.
For PBE0, closed and open symbols indicate values obtained with the
singularity correction turned on and off, respectively.
Arrows show data points which were also obtained
with $\Gamma$-point sampling. The energies obtained with
the two functionals are aligned through the local
electrostatic potential (cf.\ Ref.\ \onlinecite{Alkauskas_PRL_2008}).
}
\end{figure}

Next, we addressed the convergence of energy band edges and band gaps.
The band gap is obtained from the energy eigenvalues:
\begin{equation}
E_{\text{g}} = \varepsilon_{\text{c}} -\varepsilon_{\text{v}}
\label{BG1}
\end{equation}
where $\varepsilon_{\text{c}}$ and $\varepsilon_{\text{v}}$ are
the conduction band minimum and the valence band maximum, respectively.
In Fig.~\ref{fig5},
we give the band gaps of silicon and SiO$_2$ as a function of $k$-point
sampling or supercell size, as obtained within both the PBE and the PBE0.
The results obtained with the hybrid functional are given with and without
the singularity correction. In analogy with the convergence of the total
energy (Fig.\ \ref{fig4}), the convergence of the band gap is very
slow when omitting this correction.
The improvement is more clear for SiO$_2$, for which the band gap
converges at a sparser $k$-point density. Results obtained with large
supercells and $\Gamma$-point sampling coincide with those obtained
with small unit cells and dense $k$-point meshes. We note that in the
case of SiO$_2$ a converged value of the PBE0 band gap is already
achieved with a simulation cell of 72 atoms, well within current
computationally accessible limits. In the case of Si, the convergence
is slower because of the smaller band gap, but the convergence of
the hybrid functional result is similar to that achieved with the
semilocal functional when the singularity correction is included.
%
%
%
In Fig.~\ref{fig5}, the convergence of the respective band edges is shown.
The singularity correction only affects the occupied states by introducing a
constant downward shift of the energies (cf.\ Sec.\ \ref{Divergence}).

\begin{figure}
\includegraphics[width=8.5cm]{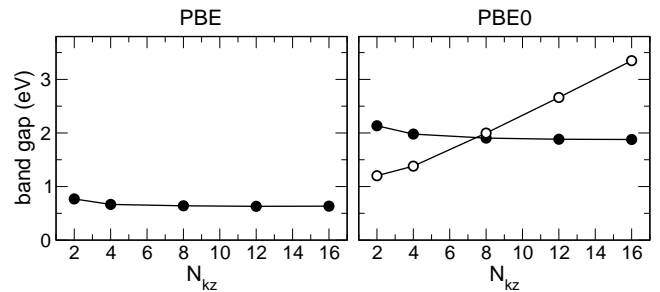}
\caption{ \label{fig6} Silicon band gap calculated with the PBE (left panel)
and PBE0 (right panel) functionals for a cubic simulation cell containing
8 atoms vs number of $k$-points in the [001] direction,
$N_{kz}$, with fixed $k$-point sampling in the orthogonal
directions ($N_{kx}$=$N_{ky}$=2).
For PBE0, closed and open symbols indicate values obtained with the
singularity correction turned on and off, respectively.}
\end{figure}

In many applications concerning surfaces and interfaces, the supercells
have an elongated shape to describe the transition across the boundary
region.  For such systems, the omission of the singularity
correction in a hybrid functional calculation in which a $\Gamma$-point
sampling is used leads to a peculiar behavior of total energies and
single particle eigenvalues. To simulate these conditions, we considered
a cubic 8-atom simulation for bulk silicon and increased the $k$-point sampling
along the [001] direction, while keeping the $k$-point sampling in the
orthogonal directions constant. This description is equivalent to that
achieved with an elongated supercell calculation in which the Brillouin zone
is sampled at the sole $\Gamma$ point. Figure \ref{fig6} shows the
calculated band gap vs.\ the number of $k$-points in the [001] direction.
Omitting the singularity correction leads in this case to a linear
increase of the band gap. A similar linear increase is also found for the
total energy (not shown). In neither case, it is therefore possible to
obtain a converged value. Including the singularity correction reestablishes
a converging behavior that resembles that found in semilocal density
functional calculations.

From the result in Fig.\ \ref{fig6}, we infer that the singularity corrections
for elongated simulation cells may change sign in comparison with cubic unit
cells and/or isotropic $k$-point meshes. To understand this behavior, it is
useful to recall that, in the case of $\Gamma$-point sampling, the singularity
correction is proportional to the energy per unit cell of a periodically
repeated point charge immersed in a compensating background.
In case of nearly cubic supercells the latter energy is negative, i.e.\ the
attractive interaction of the point charge with the uniform background is
larger than the repulsive interactions between the point charge and its images.
When the shape of the cell elongates in one or two directions, the repulsive
interaction with the image charges in the orthogonal directions grows
because of the reduced screening. For sufficiently elongated shapes, this
repulsive interaction dominates and the sign of the point-charge energy
switches.

To summarize the results of this section, we showed that the singularity
correction is needed to obtain converged total energies and single-electron
eigenvalues in hybrid functional calculations with small unit cell and dense
$k$-point samplings, in accord with previous
studies.\cite{Carrier_PRB_2007,Spencer_PRB_2008}
Furthermore, we showed that hybrid functional calculations with large
supercells and $\Gamma$-point samplings also benefit from the singularity
correction, yielding levels of convergence which are similar to those
achieved with semilocal functionals. In the case of $\Gamma$-point
samplings, the singularity correction applies to both molecular and
extended systems in a qualitatively similar way.

\section{Charged systems: Localized states} \label{Charge-Loc}

In the following, we discuss convergence issues associated to charged systems
when using hybrid functional schemes based on plane wave basis sets.
Charged systems occur in several circumstances, such as, for example, when
studying charged molecules, defects, ions in liquids, etc. We here focus on
the determination of total energy differences between different charge states
and the way the singularity correction affects the convergence properties.
We are particularly interested in assessing how the convergence properties
of hybrid functional calculations compare with those of semilocal density
functional calculations. For simplicity, we consider from now on only systems
in which the electronic structure is sampled through the sole $\Gamma$ point.
The convergence is therefore studied with respect to increasing simulation
cell, or equivalently with respect to decreasing singularity correction.
Generalization to convergence with $k$-point samplings is trivial.
In this section, we deal with atomically localized states, either in finite
systems or as defect states in solids. Infinitely delocalized states are
discussed in Sec.\ \ref{Charge-Deloc}. In this work, we do not
consider states showing intermediate degrees of localization.
For a discussion of the latter, we defer the reader to Refs.\
\onlinecite{Ogut_PRL_1998}.

\subsection{Finite systems} \label{finite}

To illustrate the convergence of total energy differences between different
charge states, we considered the calculation of the ionization potential
(IP) of an isolated molecule:
\begin{equation}
\text{IP}=E_{N-1}-E_{N},
\label{IP}
\end{equation}
where $E_{N}$ is the total energy of the neutral molecule and $E_{N-1}$ the
total energy of the same molecule in which one electron has been removed.
The total energy calculations correspond to an isolated molecule placed
in a large supercell and subject to periodic boundary conditions. This
technical constraint introduces a spurious interaction between the localized
charge and the neutralizing background charge, which needs to be considered
in order to speed up the convergence.\cite{Makov_PRB_1995} This correction
depends on the size of the simulation cell, but applies indifferently to both
hybrid functionals and semilocal density functionals. In our calculations,
the dominant correction corresponding to the charge monopole has systematically been included.\cite{Makov_PRB_1995}

\begin{figure}
\includegraphics[width=8.5cm]{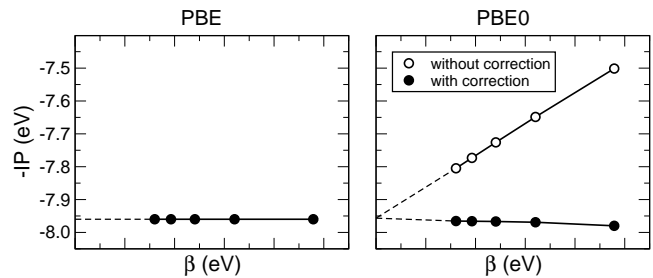}
\caption{Ionization potential (IP) of naphthalene calculated with the PBE (left panel)
and PBE0 (right panel) functionals for cubic simulation cells vs singularity
correction $\beta$, which scales like the inverse of the simulation cell size.
For PBE0, closed and open symbols indicate values obtained with the
singularity correction turned on and off, respectively.}
\label{fig7}
\end{figure}

In Fig.\ \ref{fig7}, we give the ionization potentials of naphthalene
calculated within both the PBE and the PBE0 for different supercell sizes.
The PBE0 results are obtained by both including and dismissing the singularity
correction. The results are plotted as a function of the singularity
correction, which scales like the inverse of the simulation cell size.
For the simulation cells considered (cubic cells with sides ranging between 20
and 60 bohr), the PBE results are already close to the converged values obtained
by linear extrapolation. The same consideration applies for the PBE0 results
which include the singularity correction. However, in absence of singularity
correction, the error with respect to the converged values is significantly
larger.  The converged PBE0 values for the ionization potentials of naphthalene
and pyridine are reported in Table \ref{tab1}, where they are compared with
results obtained with an all-electron scheme based on localized orbitals.
The values calculated within the two schemes differ by less than 0.05 eV.

It is of interest to extend our comparative study between PBE and PBE0
calculations to the approximate scheme based on Slater's transition
state.\cite{Slater_AQC_1972}  According to the integral form of
Janak's theorem,\cite{Janak_PRB_1978} the total-energy difference of
Eq.\ (\ref{IP}) is given by
\begin{equation}
E_{N-1}-E_{N}= -\int_0^1 \varepsilon_n(f)df,
\label{Janak1}
\end{equation}
where $\varepsilon_n(f)$ describes the eigenvalue of the highest
occupied eigenstate $n$ as it varies with its fractional occupation $f$.
Using the trapezoidal rule for the integral, we obtain Slater's approximation:
\begin{equation}
E_{N-1}-E_{N}\approx -\varepsilon_n({1}/{2})
             \approx -\left[\varepsilon_n(0)+\varepsilon_n(1)\right]/2,
\label{Janak2}
\end{equation}
where we further approximated the eigenvalue at half-filling by an average
at integer occupations. For the molecules investigated here, this approximation
is found to give accurate results for both the semilocal and the hybrid
functional (not shown). Note that, in the latter case, the singularity
corrections of energies and eigenvalues are compatible with the relation in
Eq.\ (\ref{Janak2}).  Indeed, this approximation equally holds for corrected
energies and eigenvalues as for uncorrected ones. Similar considerations
apply to electron removal energies.

\subsection{Defects in solids} \label{Defects}

The discussion pertaining to total energy differences of finite systems (Sec.\
\ref{finite}) applies with minor modifications to the study of localized defect
states in solids. In this case, the relevant physical quantities, the charge
transition levels, are also expressed as total energy differences.
Defect formation energies are first determined for varying electron chemical
potential $\mu$:\cite{VanDeWalle_JAP_2004}
\begin{equation}
E_{\text{f}}^{q}(\mu)=E^{q}_{\text{tot}}-E^{\text{bulk}}_{\text{tot}}
-\sum_{\alpha}n_{\alpha}\eta_{\alpha}+q(\mu+\varepsilon_\text{v} + \Delta V)
+ E_\text{corr}^q,
\label{formation}
\end{equation}
where $E^{q}_{\text{tot}}$ is the total energy of the defect system carrying
a charge $q$, $E^{\text{bulk}}_{\text{tot}}$ the total energy of the
unperturbed system, $n_{\alpha}$ the number of extra atoms of species
$\alpha$ needed to create the defect, and $\eta_{\alpha}$ the corresponding
atomic chemical potential.  The chemical potential $\mu$ is referred to the
valence band maximum $\varepsilon_\text{v}$.  $\Delta V$ is a correction
which is applied in order to align the local potential far away from the
neutral defect to that of the unperturbed bulk.\cite{VanDeWalle_JAP_2004}
The correction $E_\text{corr}^q$ describes the spurious interaction of the
added charge with the compensating background charge.\cite{Makov_PRB_1995}
As commented in Sec.\ \ref{exchange}, the leading term of this correction
pertaining to the monopole can be expressed as
\begin{equation}
E_\text{corr}^q=\frac{q^2\chi}{2\epsilon},
\end{equation}
where $\chi$ is defined in Eq.\ (\ref{beta3}) and $\epsilon$ is the
dielectric constant of the unperturbed bulk system. In our calculations,
this correction is applied systematically in both PBE and PBE0 calculations.

Charge transition levels correspond to specific values of the electron
chemical potential for which two charge states have equal formation energies.
For example, the charge transition level $\mu_{q/q'}$ between two charge states
$q$ and $q'$ is defined by the condition $E_{\text{f}}^{q}=E_{\text{f}}^{q'}$ and is thus
given by the following expression:
\begin{equation} \label{muu}
\mu_{q/q'} = \frac{\left(E_\text{\text{f}}^{q'}-E_\text{\text{f}}^{q}\right)
+\left(E_\text{corr}^{q'}-E_\text{corr}^{q}\right)}
{q-q'} - (\varepsilon_\text{v} + \Delta V).
\end{equation}
In this expression the dependence on the atomic chemical potentials $\eta$ drops
out and the defect charge transition level is basically determined by a
total energy difference between different charge states of the defect.
In this sense, these quantities are counterparts of the ionization potentials
and electron affinities of isolated atomic and molecular systems.
The charge transition levels of localized defect states can also be obtained
in a very accurate way through the energy eigenvalues by the application
of Janak's theorem both in PBE and in PBE0 [Eq.\ (\ref{Janak2})].\cite{Alkauskas_PhB_2007b}

\begin{figure}[t]
\includegraphics[width=8.5cm]{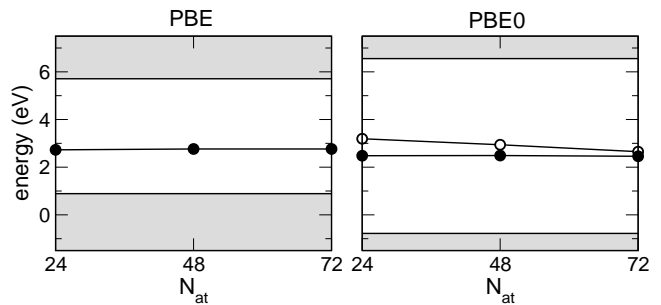}
\caption{\label{fig8} Vertical charge transition levels $\mu_{+/0}$
associated to the hydrogen bridge defect in $\beta$-crystobalite as
obtained with the PBE (left panel) and the PBE0 (right panel) functionals
vs number of atoms $N_\text{at}$ in the simulation cell.
For PBE0, closed and open symbols indicate values obtained with the
singularity correction turned on and off, respectively.
The indicated valence and conduction band edges correspond to the
converged energy eigenvalues. The energies
obtained with the two functionals are aligned through the local
electrostatic potential (cf.\ Ref.\ \onlinecite{Alkauskas_PRL_2008}).}
\end{figure}

We illustrate the convergence behavior of charge transition levels for
the hydrogen bridge defect (hydrogen substitutional to oxygen) in
$\beta$-crystobalite SiO$_2$. In particular, we focused on the vertical
charge transition level $\mu_{+/0}$ located at approximately mid
gap.\cite{Bloechl_PRB_2000,Alkauskas_PhB_2007}
We considered three different supercells in which one lateral size was varied
while the other dimensions were kept fixed, i.e.\ we used
2$\times$2$\times$2, 2$\times$2$\times$3, and 2$\times$2$\times$4 unit cells.
Such elongated cells might for instance occur when slab models are adopted.
Charge transition levels calculated in the PBE and in the PBE0 are shown
in Fig.~\ref{fig8} for increasing number of atoms in the simulation cell.
The results obtained with the two functionals are aligned through
the local electrostatic potential, as suggested by the study in Ref.\
\onlinecite{Alkauskas_PRL_2008}.
It clearly appears that the PBE and PBE0 defect levels show a similar
convergence behavior, provided the singularity correction is included in the
PBE0 calculation. The singularity correction is crucial to achieve this level
of convergence. Without the singularity correction, the charge transition
levels are clearly not converged for the range of simulation cells considered.
It is clearly seen that the convergence behavior of charge transition levels
of defects is analogous to that of energy transitions in finite systems
(Sec.\ \ref{finite}).

\section{Charged systems: Delocalized states} \label{Charge-Deloc}

The case in which the extra charge is carried by extended delocalized states
requires special attention. Let us assume an infinite solid with a energy
eigenvalue spectrum $\varepsilon_{k}$. The energy cost of adding one electron
to the previously unoccupied state $n$ is simply given by
\begin{equation}
E_{N+1}-E_{N}=\varepsilon_n,
\label{Janak3}
\end{equation}
where we used Janak's theorem\cite{Janak_PRB_1978} as formulated in Eq.\
(\ref{Janak1}) and the fact that energy eigenvalues in infinite solids
do not depend on the occupation of the state. Identical considerations
also apply for electron removal energies. On this basis, it is possible
to express the valence and conduction band edges in terms of total energy
differences between systems of different charge:
\begin{eqnarray}
\tilde{\varepsilon}_{\text{c}}&=&E_{N+1}-E_{N}, \label{Janak4} \\
\tilde{\varepsilon}_{\text{v}}&=&E_{N}-E_{N-1}. \label{Janak5}
\end{eqnarray}
Consequently, the energy band gap can also be obtained as
\begin{equation}
\tilde{E}_\text{g}=
\tilde{\varepsilon}_{\text{c}} -\tilde{\varepsilon}_{\text{v}}=E_{N+1}+E_{N-1}-2E_{N}.
\end{equation}
In our notation, the tilde sign signifies that the concerned quantity
is obtained from a total energy difference, as opposed to
a direct derivation from the spectrum of energy eigenvalues.

In practical calculations, the supercells always have {\it finite} size,
by which the band edges determined by total-energy difference differ
from the actual energy eigenvalues:\cite{Lany_PRB_2008}
\begin{eqnarray}
\tilde\varepsilon_{\text{c}}&=&{\varepsilon}_{\text{c}} + \Delta\varepsilon_\text{c}, \label{Janak6}\\
\tilde\varepsilon_{\text{v}}&=&{\varepsilon}_{\text{v}} + \Delta\varepsilon_\text{v}, \label{Janak7}
\end{eqnarray}
where it is understood that
\begin{eqnarray}
\lim_{\Omega\rightarrow \infty} \Delta\varepsilon_\text{c} &=& 0, \\
\lim_{\Omega\rightarrow \infty} \Delta\varepsilon_\text{v} &=& 0,
\end{eqnarray}
where $\Omega$ is the volume of the simulation cell. Similarly, we define
\begin{equation}
\tilde{E}_{\text{g}}={E}_{\text{g}} + \Delta E_{\text{g}},
\end{equation}
where $\Delta E_{\text{g}}=\Delta\varepsilon_\text{c}-\Delta\varepsilon_\text{v}$.

\begin{figure}
\includegraphics[width=8.5cm]{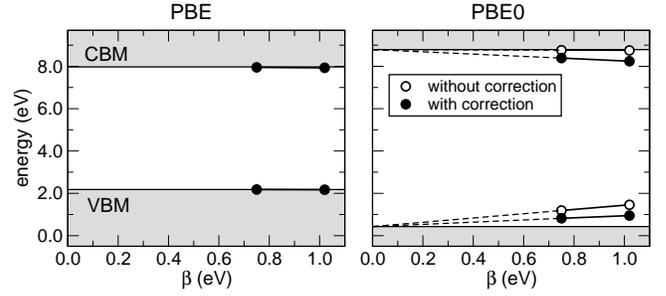}
\caption{Band edges of $\alpha$-quartz calculated with the PBE (left)
and PBE0 (right) functionals as total energy differences vs singularity
correction $\beta$ that characterizes the simulation cell, the limit
of infinite simulation cell being achieved for vanishing $\beta$.
For PBE0, closed and open symbols indicate values obtained with the
singularity correction turned on and off, respectively. The energies
obtained with the functionals are aligned as in Fig.\ \ref{fig5}.}
\label{fig9}
\end{figure}

We first focus on results obtained with the semilocal density functional.
For illustration, we considered $\alpha$-quartz SiO$_2$ and used simulation
cells of varying size. In Fig.\ \ref{fig9}, we report the band edges calculated
using the total energy differences given in Eqs.\ (\ref{Janak4})
and (\ref{Janak5}), and compare them with the converged energy eigenvalues.
For the simulation cells considered here, the band edges obtained in the
two different ways are essentially identical, i.e.\
$\Delta\varepsilon_\text{c}\approx 0$ and $\Delta\varepsilon_\text{v}\approx 0$.

Figure \ref{fig9} also shows corresponding results obtained with hybrid
functionals. In contrast to the behavior found for semilocal functionals,
we now encounter a much slower convergence. In other words, in hybrid
functional calculations, $\Delta\varepsilon_\text{c}$ and
$\Delta\varepsilon_\text{v}$ are significantly larger than in semilocal
density functional calculations performed with the same simulation cells.
Therefore, we infer that this behavior should be ascribed to the
singular behavior of the exchange interaction.

\begin{figure}
\includegraphics[width=8.5cm]{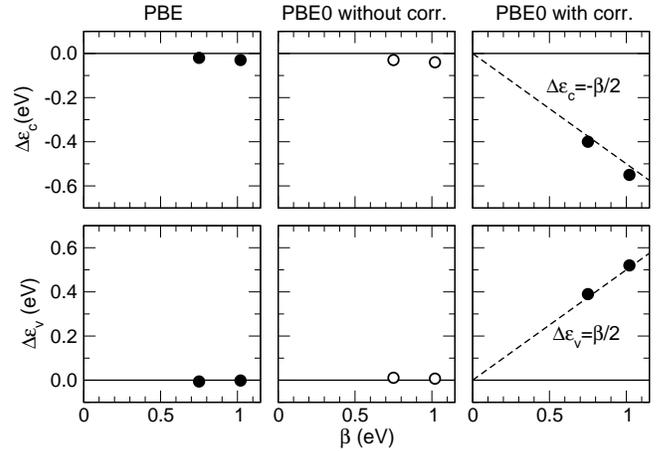}
\caption{Difference between band edges of $\alpha$-quartz evaluated through
total energy differences
($\tilde{\varepsilon}_{\rm c}$, $\tilde{\varepsilon}_{\rm v}$) and as energy
eigenvalues (${\varepsilon}_{\rm c}$, ${\varepsilon}_{\rm v}$)
in the PBE (left panel), (b) in the PBE0 with singularity turned off (middle panel),
and (c) in the PBE0 with the singularity correction turned on (right panel). The
results are plotted against the singularity correction $\beta$.
In the right panel, the dashed lines correspond to
$\Delta \varepsilon_{\rm c}=-\beta/2$ and $\Delta \varepsilon_{\rm v}=+\beta/2$.}
\label{fig10}
\end{figure}

In order to understand this behavior, we reason as follows.
The relations given by Eqs.\ (\ref{Janak6}) and (\ref{Janak7})
follow from the Janak theorem and apply to any analytical functional.
Therefore, they also apply to a case functional which is specific
to a given simulation cell and a given basis set. We further define the
case functional by setting the $\mathbf{G}=0$ component of the exchange
potential to zero, which corresponds to the label ``uncorr'' introduced
earlier.  For this case functional, $\Delta\varepsilon_\text{c}$ and
$\Delta\varepsilon_\text{v}$ are expected to behave in a similar way
as for a semilocal functional. Hence, for simulation cells large enough
to yield vanishing $\Delta\varepsilon_\text{c}$ and
$\Delta\varepsilon_\text{v}$ with the semilocal functional, we similarly
expect vanishing $\Delta\varepsilon^\text{uncorr}_\text{c}$ and
$\Delta\varepsilon^\text{uncorr}_\text{v}$. This implies
\begin{eqnarray}
\varepsilon^\text{uncorr}_{\text{c}}&\approx&E^\text{uncorr}_{N+1}-E^\text{uncorr}_{N} , \\
\varepsilon^\text{uncorr}_{\text{v}}&\approx&E^\text{uncorr}_{N}-E^\text{uncorr}_{N-1} .
\end{eqnarray}
As shown in the first two panels in Fig.\ \ref{fig10}, this behavior
is numerically confirmed for our case study of $\alpha$-quartz.

Using the relationship between corrected and uncorrected quantities
determined in Eqs.\ (\ref{corr_eigen}) and (\ref{corr_en}), we then derive:
\begin{eqnarray}
\varepsilon^\text{corr}_{\text{c}}&\approx&E^\text{corr}_{N+1}-E^\text{corr}_{N} + \beta/2, \\
\varepsilon^\text{corr}_{\text{v}}&\approx&E^\text{corr}_{N}-E^\text{corr}_{N-1} - \beta/2.
\end{eqnarray}
In other words, hybrid functional calculations in finite simulation cells give
\begin{eqnarray}
\Delta\varepsilon_\text{c} &\approx &- \beta/2,\\
\Delta\varepsilon_\text{v} &\approx &+ \beta/2.
\end{eqnarray}
This result is graphically illustrated for $\alpha$-quartz in the third-column
panels of Fig.\ \ref{fig10}.  For the band gap, this implies:
\begin{equation}
\Delta E_{\text{g}}\approx - \beta.
\end{equation}

We note that the remaining dependence of both $\tilde{\varepsilon}_\text{c}$
and $\tilde{\varepsilon}_\text{v}$ on $\beta$ does not result from the
integration of the singularity, since the results in the third panel
in Fig.\ \ref{fig10} already refer to ``corrected'' results. The remaining
differences rather originate from the exchange selfinteraction associated
to the extra charge, which vanishes slowly with increasing simulation cell.
Hence, this behavior is not specific to the use of plane-wave basis sets
and should also manifest in implementations based on other basis functions.

\begin{figure}
\includegraphics[width=8.5cm]{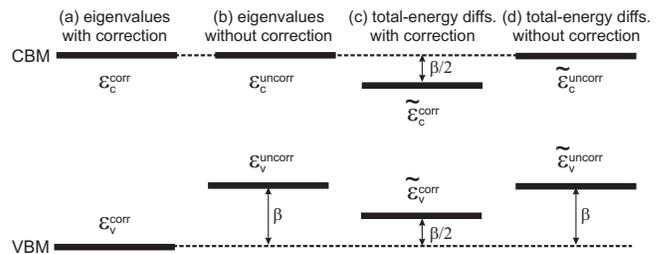}
\caption{Schematic representation of the band edges determined with hybrid
functionals in the presence of a limited $k$ point sampling: as energy
eigenvalues (a) with and (b) without accounting for the singularity correction
and through total energy differences (c) with and (d) without singularity
correction. The result in (a) shows the fastest convergence with simulation cell.}
\label{fig11}
\end{figure}

Our findings concerning the energy eigenvalues are schematically illustrated
in Fig.\ \ref{fig11}. The scheme refers to a case in which all quantities would
already be converged with simulation cell size, if it were not for the
occurrence of exact exchange. Figure \ref{fig11}(a) refers to the corrected
energy eigenvalues and corresponds to the converged result. Figure
\ref{fig11}(b) shows the eigenvalues in the ``uncorrected'' case.
The singularity correction shifts the occupied states downwards by $\beta$,
leaving the unoccupied ones unaffected. Figure \ref{fig11}(c) refers to
the determination of band edges through the use of total-energy differences.
It is seen that the band edges obtained in this way still differ from the
converged levels, despite the use of ``corrected'' quantities. The valence
band edge is overestimated by $\beta/2$, whereas the conduction band edge
is underestimated by $\beta/2$. Consequently, the band gap is underestimated
by $\beta$ through this approach. Figure \ref{fig11}(d) also refers to band
edges obtained through total energy differences but without including the
singularity correction. One obtains the same result as in Fig.\ \ref{fig11}(b),
illustrating thereby that the integration of the singularity in the
exchange term is responsible for the slow convergence of the band edges
calculated by total-energy differences.

\section{Conclusions \label{Conclusions}}

In this work, we investigated the use of hybrid functionals in
plane-wave implementations, in comparison with semilocal density functionals.
The main objective consisted in determining whether a hybrid
functional calculation requires a $k$-point sampling of increased
density in order to properly integrate the singularity appearing in the
exact nonlocal exchange energy. This issue is of particular importance
when dealing with large simulation cells, where an excessive increase
of the $k$-point sampling would make the calculation computationally
prohibitive. Typical applications include surface, interface, and
defect calculations, but also molecular dynamics simulations.

To treat the divergence, we adopted a formulation which consists in
transforming the integrand into a regular function through the use
of an auxiliary function that can be integrated
analytically.\cite{Gygi_PRB_1986} Through the use of an appropriate
auxiliary function,\cite{Massidda_PRB_1993} this formulation can
trivially be extended to calculations with large simulation
cells and low-density $k$-point samplings. In the case of $\Gamma$-point
sampling, the sampling of reciprocal space is achieved through the
reciprocal lattice vectors, which densify as the simulation cell grows.
We further used a formulation which recasts the treatment of the
divergence in the form of singularity correction terms.\cite{Carrier_PRB_2007}
These terms intervene in the total energy and in the energy eigenvalues
of the occupied states.

In the present investigation, we found it convenient to distinguish
finite and extended systems, localized and delocalized states, and
neutral and charged calculations. The general conclusion is that
the same $k$-point samplings used in semilocal density functional
calculations yield a comparable level of convergence in hybrid
functional calculations, provided the singularity corrections are
accounted for. However, our study highlights a few points that deserve
special attention.

The first point concerns the treatment of screened exchange. While
screened exchange does not show any singularity, it is nevertheless
recommended to adopt the proposed scheme also in this case in order
to achieve the same convergence properties as achieved with semilocal
density functionals.

The second noteworthy point concerns applications in which the
sampling in reciprocal space around the divergence is anisotropic.
This is for instance the case for calculations with elongated
simulation cells and $\Gamma$-point samplings, as often occurs
in the study of surfaces and interfaces. In such cases,
the singularity correction terms are critical for achieving not only
well converged properties but also a qualitatively correct behavior.

The final point that deserves attention and which is unusual
with respect to ordinary semilocal density functional calculations
concerns the determination of band edges and band gaps of extended
systems through total energy differences. Our study shows that
band edges determined in this way converge slower in hybrid
functional calculations because of the exchange selfinteraction
associated to the extra charge. This convergence problem arises
for delocalized states but does not occur for localized states
of point defects or of finite molecular systems.

In conclusion, the correct treatment of the singularity can be achieved
without requiring any significant computational overhead.
This opens the way for using hybrid functionals in very much the
same way as ordinary semilocal functionals.
To date, the scheme described in this work has already led to
several successful applications including studies of
amorphous systems,\cite{Broqvist_APL_2007}
defects,\cite{Broqvist_APL_2006,Broqvist_APL_2008,Alkauskas_PRL_2008,%
Broqvist_PRB_2008,Alkauskas_PRB_2008,Broqvist_JAP_2009}
and interfaces.\cite{Devynck_PRB_2007,Devynck_APL_2007,Broqvist_APL_2008,%
Alkauskas_PRL_2008b,Devynck_SS_2008,Broqvist_APL_2009}

\section*{Acknowledgments}
We thank S.~de Gironcoli and J.~Hutter for fruitful interactions.
Support is acknowledged from the Swiss National Science Foundation
(Grants Nos.\ 200020-111747 and 200020-119733). Calculations
were performed on the BlueGene computer at EPFL and on computer
facilities at CSEA-EPFL and CSCS.

\end{document}